\documentclass[amsmath, amsfonts, superscriptaddress, showpacs, twocolumn, prl]{revtex4}
\usepackage{graphicx}
\usepackage{epsfig}
\usepackage{bm}
\usepackage{dcolumn}
\usepackage{amsmath}
\usepackage{amssymb}

\begin{document}

\title{Distribution of supercurrent switching in graphene under proximity effect}

\author{U.~C.~Coskun}
\email[]{uccoskun@gmail.com}
\affiliation{Department of Physics, University of Illinois at
Urbana-Champaign, Urbana, Illinois 61801, USA}
\affiliation{Department of Physics, University of Texas, Dallas,
Texas 75080, USA}

\author{M.~Brenner}
\affiliation{Department of Physics, University of Illinois at
Urbana-Champaign, Urbana, Illinois 61801, USA}

\author{T.~Hymel}
\affiliation{Department of Physics, University of Illinois at
Urbana-Champaign, Urbana, Illinois 61801, USA}

\author{V.~Vakaryuk}
\affiliation{Materials Science Division, Argonne National
Laboratory, Argonne, Illinois 60439, USA}

\author{A.~Levchenko}
\affiliation{Department of Physics and Astronomy, Michigan State
University, East Lansing, Michigan 48824, USA}

\author{A.~Bezryadin}
\affiliation{Department of Physics, University of Illinois at
Urbana-Champaign, Urbana, Illinois 61801, USA}

\begin{abstract}
We study the stochastic nature of switching current in hysteretic current-voltage
characteristics of superconductor-graphene-superconductor (SGS) junctions. We find
that the dispersion of the switching current distribution scales
with temperature as $\sigma_I\propto T^{\alpha_{G}}$ with
$\alpha_{G}$ as low as $1/3$. This observation is in sharp contrast with
the known Josephson junction behavior where $\sigma_I\propto
T^{\alpha_{J}}$ with $\alpha_{J} =2/3$. We propose an explanation
using a generalized version of Kurkij\"arvi's theory for the flux
stability in rf-SQUID and attribute this anomalous effect to the
temperature dependence of the critical current which persists down
to low temperatures.
\end{abstract}

\date{December 2, 2011}

\pacs{74.45.+c, 72.80.Vp, 74.40.-n, 74.50.+r}

\maketitle

Since the extraction of single-layer graphene~\cite{Novoselov-05,Zhang-05} much effort has
concentrated on its study due to the promising potential in
applications. The knowledge of graphene properties and expertise in
making high quality devices have grown
substantially~\cite{Geim-07,Geim-09,Castro-Neto-09}. Nevertheless,
the transport in graphene subject to nonequilibrium conditions and
in the proximity to a superconductor, an important ingredient in the
majority of applications, is far from being fully understood. Unlike
metal-superconductor interfaces reflection from a
graphene-superconductor boundary is governed by the
\textit{specular} Andreev processes~\cite{Beenakker}. This peculiar
effect combined with the unique band structure of graphene makes
proximity effect in graphene a particularly interesting subject to
study.

Recent experiments on the superconductor-graphene-superconductor
(SGS) devices have revealed many interesting features caused by the
proximity effect~\cite{Meissner-PRL59}. These include an observation
of supercurrent and subsequent measurement of the current-phase
relation, signatures of multiple Andreev reflection in the
differential conductance, and Shapiro steps  under microwave
irradiation, see Refs.~\cite{Heersche-Nature07,
Miao-Science07,Du-PRB08,Ojeda-PRB09,Jeong-PRB11}. Recent
measurements have also revealed the residual resistance of SGS
junctions for currents below critical which was attributed to the
phase diffusion phenomenon \cite{Borzenets} followed by the
crossover to macroscopic quantum tunneling regime at low
temperatures~\cite{Lee-arXiv11}. Here we report first systematic
study of thermally activated dynamics of phase slips  in SGS
junctions through the measurement of the switching current
distribution.

Measurement of the decay statistics of metastable states is a
powerful tool for revealing the intrinsic thermal and quantum
fluctuations. In the Josephson junctions (JJ) a metastable
dissipationless (superconducting) state decays into dissipative
(phase slippage) state when the bias current $I$ reaches a critical
value called switching current $I_{SW}$, which is stochastic.
Analysis of the distribution of the switching current was employed
to reveal macroscopic quantum tunneling in JJs~\cite{MDC},
superconducting nanowires~\cite{Sahu}, small underdamped
JJs~\cite{Yu} and intrinsic JJs in high-$T_c$
compounds~\cite{Warburton2009}. Experimentally observed temperature
dependence of the switching current dispersion $\sigma_I$ always
follows a power law $\sigma_I\propto T^{2/3}$, if the switching is
induced by a single thermally activated phase slip~\cite{FD}.
However at sufficiently low temperatures the temperature dependence
of $\sigma_I$ saturates, which is usually attributed to the
macroscopic quantum tunneling~\cite{MDC}.

Study of switching current distribution in conventional SNS
junctions, where N is normal metal, is obstructed by the fact that
such junctions are usually overdamped. As a result  their $I$-$V$
characteristics are smooth and the notion of the switching current
is not applicable. Here we report a study of moderately underdamped
SGS junctions with the quality factor $Q\simeq 4$ for the entire
span of gate voltages. Our main finding is the anomalous temperature
dependence of the switching current dispersion $\sigma_I\propto
T^{\alpha_{G}}$ in SGS devices with
$0.3\lesssim\alpha_{G}\lesssim0.5$, which persists for a wide range
of gate-induced doping and is significantly smaller than the usual
$\alpha_{J} =2/3$. In general, any power law different from $2/3$ is
associated with the possibility of quantum phase slips. In our
graphene-based proximity junctions, although the power law notably
deviates from $2/3$, we argue that thermally activated phase slips
are the major contributor. We interpret an anomalous dispersion of
$\sigma_I$ by using a generalized Kurkij\"{a}rvi model~\cite{JK}.
Our conclusion is that the slowed temperature scaling of $\sigma_I$
in SGS junction is due to the substantial temperature dependence of
the critical current, which persists down to low temperatures in SGS
systems.

\begin{figure}[t!]
  \includegraphics[width=8.8cm]{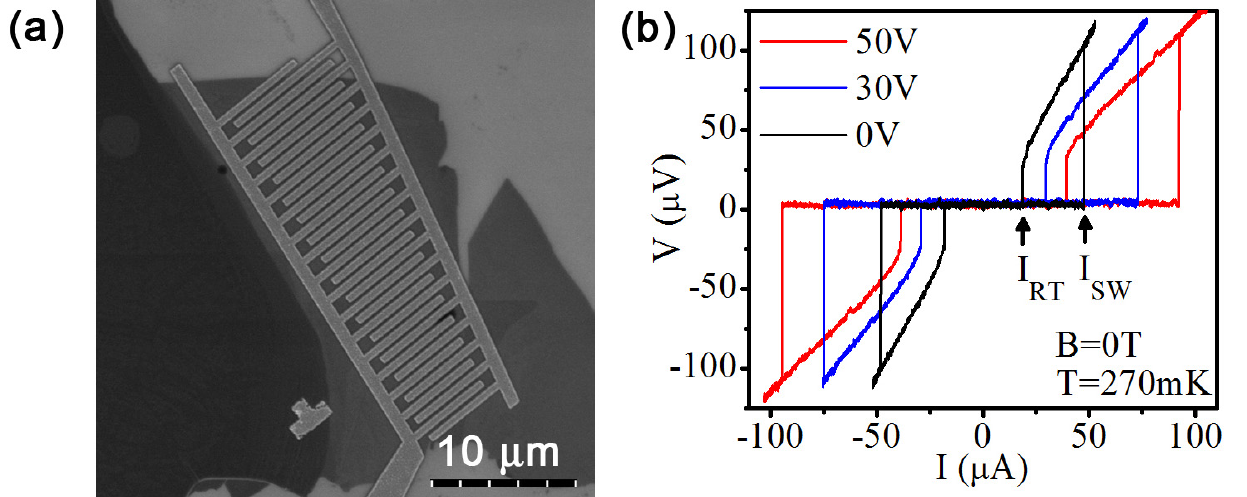}\\
  \caption{[Color online] a) SEM micrograph of sample 105.
  Distance between the electrodes along the current (length of the junction) is
  $L=265$~nm. Width of the junction (distance across the current)
  is $W=214~\mu$m. For sample 111s, $L=280$~nm and $W=9.9~\mu$m.
  b) The hysteretic $I$-$V$ curves of SGS junction (sample 105) taken at various gate voltages. The
  switching $I_{SW}$ and retrapping $I_{RT}$ currents are shown. }\label{Fig1}\vskip-.5cm
\end{figure}

Graphene flakes are deposited on 280 nm thick SiO$_2$ surface using
mechanical exfoliation~\cite{Novoselov-05}. Raman spectroscopy is
used to confirm the number of layers~\cite{Raman2}. The electrodes,
which have a fingered shape (Fig.~\ref{Fig1}a), are patterned from a
bilayer Pd/Pb (4nm/100nm), as explained in the supplementary
materials (SM). In order to measure the switching current
distribution, the amplitude of the sinusoidal current bias is set
somewhat higher than the maximum switching current, and it is
adjusted when needed to keep the sweep speed roughly constant. The
number of switching events for each distribution was either 5000
(for the sample 111s) or 10000 (for the sample 105). At low
temperatures the $I$-$V$ curves of the samples exhibit a hysteretic
behavior, Fig.~\ref{Fig1}b, which enables us to study switching
current statistics.

\begin{figure}[b!]
  \includegraphics[width=8.8cm]{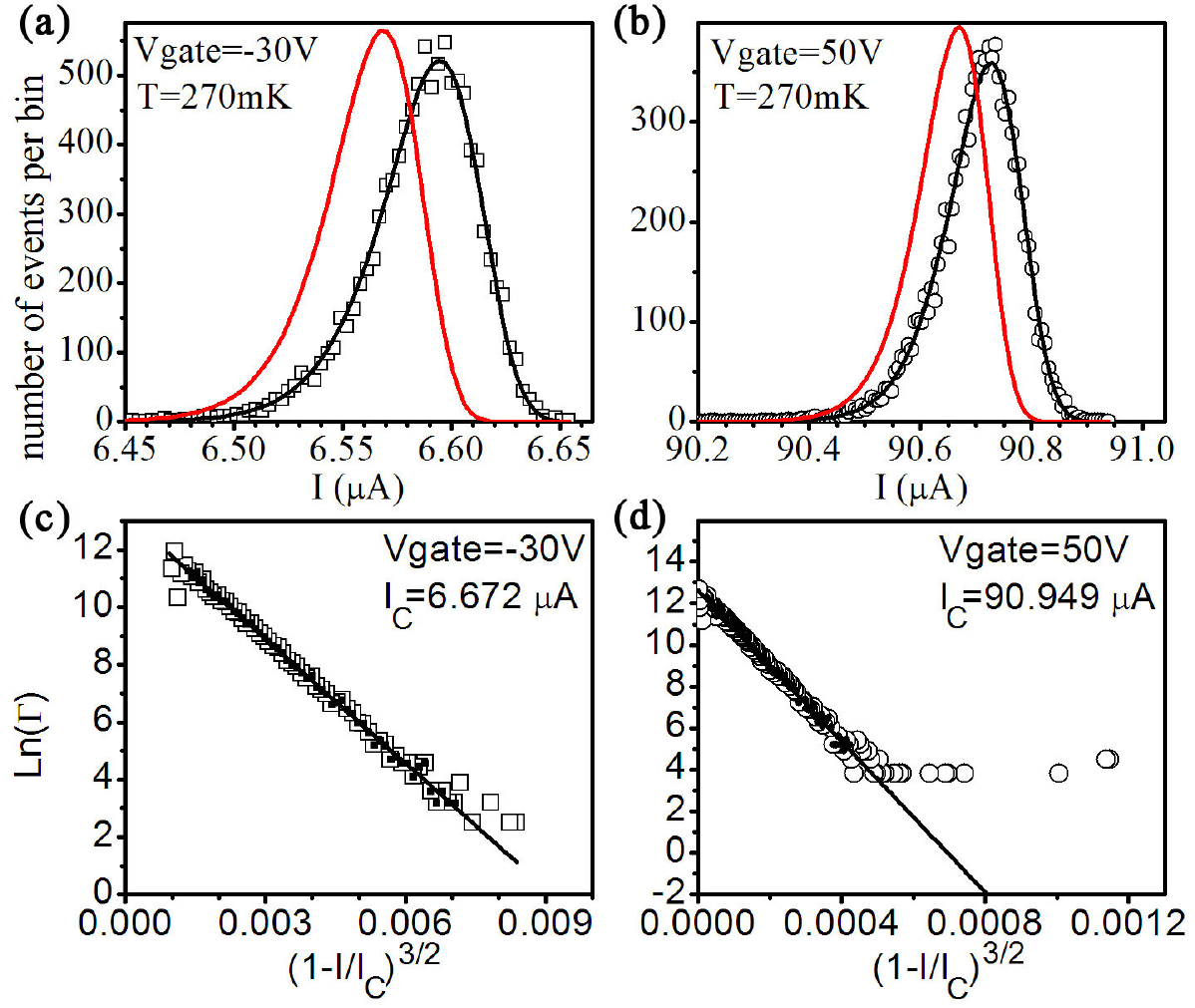}\vskip-.25cm
  \caption{
  [Color online] (a), (b) Switching current distributions at
  Dirac point ($V_g=-30\,V$) and $V_g=50\,V$. The black curve
  shows a theoretical fit to the experimental distribution with an experimental speed: (a) $363
\, \mu$A/sec and (b) $2.7\,$~mA/sec. The red curve shows a calculated
  distribution with a new sweeping speed: (a) $100\, \mu$A/sec and
 (b) $1.0\,$~mA/sec, using the escape rate to the standardized sweep
  speed. (c)--(d) Logarithm of the escape rate is
  shown as a function of the scaled current.
  The raw data are shown as hollow circles and squares. Filled squares and
  circles are used to calculate the critical current and to fit the escape rate,
  which is shown as a solid line. Anomalous premature jumps are
  visible as isolated data on the left side of the graph.
  }
  \label{Fig2}
\end{figure}

Our main focus is on the $\sigma_I(T)$ function. Figure~\ref{Fig2}
shows a switching histogram for sample 105 at (a) Dirac point
($V_g=-30V$) and (b) $V_g=50V$. During the experiment, some
anomalously premature switching events are recorded. These events,
which significantly deviate from the general population of the
distribution, are very rare and are believed to be unrelated to
thermal fluctuations. In order to exclude these anomalous jumps from
the standard deviation calculation we first convert the raw data to
the switching rate $\Gamma(I)$ according to the Kurkijarvi
method~\cite{JK,FD}. The Kramers and Stewart-McCumber theories
combined (see below) lead to the expectation that $\ln\Gamma\propto
(1-I/I_{C})^{3/2}$. In Figs.~\ref{Fig2}c and \ref{Fig2}d we plot
$\ln\Gamma$ versus $(1-I/I_{C})^{3/2}$. The critical current $I_C$
is tuned to make the graph as linear as possible. Then the linear
part of the graph is fit with a straight line. Hollow squares and
circles are the measured data. Filled symbols are those points which
were used to find the best linear fits. The best fit $\Gamma(I)$ is
then used to regenerate the distribution of $I_{SW}$ by inverting
the Kurkij\"{a}rvi transformation. The results are shown as black
curves in Fig.~\ref{Fig2}a and Fig.~\ref{Fig2}b, for the
experimental sweep rates were $363\, \mu$A/s and $2.7\,$mA/s. The
red curves are computed distributions for $dI/dt=100\, \mu$A/s and
$1.0\,$mA/s correspondingly. The dispersion $\sigma_I$ is then
computed for the same value of $dI/dt$ for all temperatures.

\begin{figure}[t!]
 \includegraphics[width=8.8cm]{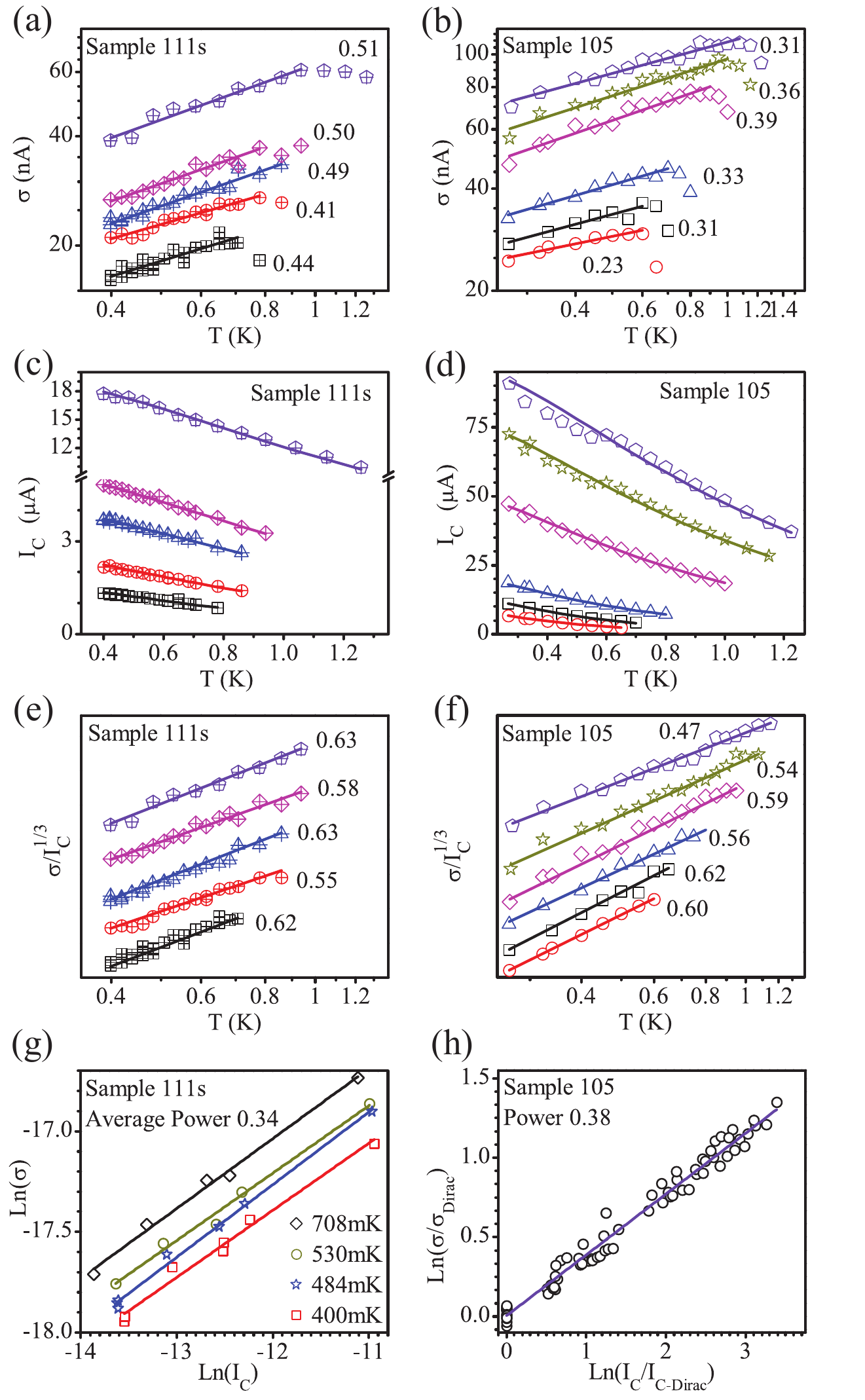}\vskip-.25cm
  \caption{[Color online] For sample 111s ($V_{gD}=-1$\,V)
 data sets on panels (a), (c) and (e) correspond to gate
 voltages $V_g$=50, 5, 3, 1, and -1~V, from top to bottom.
 For sample 105 ($V_{gD}=-30\,$V)  data sets on panels (b), (d), (f)
 correspond to $V_g$=50, 30, 10, -10, -30, and -50~V,
 from top to bottom. (a)--(b) Standard deviation vs.~temperature,
 in the log-log format. The best linear fits determine the power
 $\alpha_G$, which is shown near each fit. (c)--(d) Critical
 current vs.~temperature for sample 105 and 111s at various gate
  voltages. Solid lines are theoretical fits~\cite{ZZ}.  (e)--(f)
  Normalized standard deviation,
  $\sigma_I/I_{C}^{1/3}\propto T^{\tilde{\alpha}_G}$,  vs.~temperature,
  in the log-log format.
  The corresponding powers $\tilde{\alpha}_G$ are indicated.
  The graphs are shifted vertically for clarity. (g)
 Log-log plot of standard deviation vs.~critical current at four
 different temperatures. (h) Log-log plot of the scaled standard
 deviations vs.~scaled critical current.}
  \label{Fig3}\vskip-.5cm
\end{figure}

Our main results are presented in Fig.~\ref{Fig3}. This figure shows
standard deviation $\sigma_I$ and critical current $I_{C}$ versus
temperature for various gate voltages. Figures~\ref{Fig3}a
and~\ref{Fig3}b are log-log plots of $\sigma_I$ versus $T$. The best
linear fit provides $\alpha_G$, which is defined by the equation
$\sigma_I\propto T^{\alpha_G}$. The estimated error or uncertainty
in the power values is about 7$\%$. Overall, the best fit
$\alpha_G$'s are different from the theoretically predicted JJ value
$\alpha_{J}=2/3=0.667$. Since numerous previous experiments on JJs
established the power close to 2/3 while our data indicate powers
roughly between 1/3 and 1/2, an understanding of such discrepancy is
desirable.

We interpret these observations based on the following
model. Since the pioneering theoretical work of
Kurkij\"{a}rvi~\cite{JK} and its experimental confirmation by Fulton
and Dunkleberger~\cite{FD} kinetics of stochastic phase slips in the
JJs is described within Stewart-McCumber model~\cite{SM}, which
employs sinusoidal current-phase relation (CPR),
$I_S(\phi)=I_C\sin(\phi)$, and represents the total current as a sum
of superconducting, normal and displacement components. At the
mesoscopic scale and, in particular, in the context of graphene
proximity circuits, there are reasons to question the applicability
of such model given the possibility of a highly nontrivial structure
of $I_S(\phi)$ (see SM). This naturally raises a question
about the universality of the previous results with respect to the
form of the CPR. It is rather remarkable to realize that the
predictions of the theory~\cite{JK} in fact extend beyond the limits
of its original validity. We now proceed to the generalization of
the Kurkij\"arvi's theory \cite{JK} developed for the statistics of
thermally activated phase slips in a flux-biased rf-SQUID to the
case of a current-biased weak link with an \textit{arbitrary} CPR.

Within the Stewart-McCumber model the dynamics of the phase
$\phi$ is equivalent to the dynamics of a
viscous Brownian particle subject to the following external
potential:
\begin{equation}\label{G}
 G(\phi)= F(\phi)-\hbar I\phi/2e\,,
\end{equation}
which is the Gibbs potential. Here $ F(\phi)=(\hbar/2e)\int d\phi
I_S(\phi)$ is the free energy and $I$ is the bias current. We assume
that $I_S=I_S(\phi)$ is a single-valued smooth function. For $I=0$,
$ G(\phi)$ is a periodic function of $\phi$ with alternating local
maxima and minima. In the absence of fluctuations the phase is
trapped in one of the minima as long as $I<I_C$, which is a state
with zero voltage. In the resistive state when $I>I_C$ the phase
increases with time. In the presence of thermal fluctuations even at
$I<I_C$ the phase can escape its local minimum, i.e. experience a
phase slip, which drives the junction into a phase-running resistive
state. Such a process is detected as a voltage jump (a switching
event) on the $I$-$V$ curve. Upon decreasing $I$ the junction may
show hysteretic behavior and the quality factor $Q$ determines the
width of the hysteretic region. The activation rate of a phase slip
for a moderately underdamped ($Q \gtrsim 1$) to overdamped  ($Q\ll
1$) junction is given by Kramers theory \cite{Kramers} (hereafter
$k_B=1$)
\begin{equation}
    \Gamma
    =
    (1/2\pi)\big( \sqrt{\eta^2/4 + \omega^2} - \eta/2 \big)
    \exp(- \Delta  G/T)\,.
    \label{Kramers}
\end{equation}
The energy barrier $\Delta  G$ is the
spacing between two consecutive extrema: $\Delta
 G= G(\phi_+)- G(\phi_-)$. The
prefactor is determined by the curvature of the potential
at minimum, $\omega^2=C^{-1} (2e/\hbar)^2
\partial^2_\phi G$ and by the damping parameter  $\eta=
1/R_NC$, where $R_N$ and $C$ are effective normal resistance
and capacitance of the junction. Notice that
$Q=\omega_p/\eta$ where $\omega_p=\sqrt{2eI_C/\hbar C}$ is
plasma frequency.

To find the activation barrier $\Delta G(I)$ let us introduce a
critical phase $\phi_C$ defined through $I_C=I_S(\phi_C)$. In the
vicinity of $\phi_C$ one can use Taylor expansion
$I_S(\phi)=I_C-\frac{1}{2}|I''_C|(\phi-\phi_C)^2$ provided
$I_S(\phi)$ is a smooth function. $F(\phi)$ is obtained by
integrating the supercurrent over the phase, which gives $F(\phi)=
F_C+\frac{\hbar}{2e}I_C(\phi-\phi_C) -\frac 1 {3!}\,\frac{\hbar}{2
e}|I''_C|(\phi-\phi_C)^3$, where $ F_C= F(\phi_C)$ and
$I''_C=\partial^2_{\phi}I(\phi)|_{\phi_C}$. These equations
determine the locations of the two consecutive extrema of the Gibbs
potential: $\phi_\pm-\phi_C=\pm\sqrt{2(I_C-I)/|I''_C|}$. Using
Eq.~\eqref{G} one finds
\begin{equation}\label{G-3/2}
\Delta  G(I)= G_C(1-I/I_C)^{3/2}\,, \,\,
 G_C=\frac{2\sqrt{2}\hbar
I_C}{3e}\sqrt{\frac{I_C}{|I''_C|}}\,.
\end{equation}
The curvature of the Gibbs potential at the extrema points can be
obtained in a similar way  and is given by $\omega^2 = \omega_p^2
\sqrt{2 (|I''_C|/ I_C)(1-I/I_C)}$.

The knowledge of the decay rate (Eq.~\ref{Kramers}) allows one to
determine the probability $p$ that a phase slip occurred by the time
$t$, which reads: $p(t) = 1- e^{- \int_0^t \Gamma(t') dt'}$, where
$\Gamma=\Gamma(I(t))$. Note that the probability of not having a
phase slip by the time $t$ is $1-p$. For a constant bias current
sweep $\omega_i= I_C^{-1} d I(t)/dt$ the probability $p$ can be
evaluated analytically. Introducing reduced current variable $u=(G_C
/T)^{2/3}(1- I/I_C)$ and recalling the definition of the quality
factor one obtains the following expression:
\begin{equation}\label{p}
    p =1-e^{-X e^{-u^{3/2}}}\!\!,\quad
    X =
    \frac{2 T \omega_p^2/(3 \pi \omega_i \eta  G_C)}
    {1+ \sqrt{1+ Q^2 (T/ G_C)^{1/3} u^{1/2}}}.
\end{equation}
This result for the probability of a phase slip holds for a moderate
to high damping provided $ G_C \gg T$, the condition which is very
well satisfied in most of our measurements. In the limit of high
damping when $Q\to 0$ and $X \to T \omega_p^2/(3 \pi \omega_i \eta
G_C)$ one recovers the result of \cite{JK}.

To evaluate the dispersion of the switching current we notice that
the probability distribution $P(x)$ of a variable $x$ is obtained
from $p(x)$ by the differentiation with respect to $x$ i.e.~$P(x) =
- dp(x)/dx$. Given the relation between the bias current $I$ and the
reduced current $u$ this implies that the dispersions of these
variables are related as $\sigma_I = I_C (T/ G_C)^{2/3} \sigma_u$.
The crucial observation is that dispersion $\sigma_u$ considered as
a function of $X$ is constant within a few percent while $X$ is
varied by several orders of magnitude, so that for all practical
purposes $\sigma_u$ is temperature independent \cite{JK}. Using
Eq.~\eqref{G-3/2} and assuming that the temperature scalings of
$I_C$ and $I''_C$ are the same we obtain the following temperature
scaling for the dispersion of switching current:
\begin{equation}
    \sigma_I (T)\simeq(T/\Phi_0)^{2/3} I_C^{1/3}(T)\,,
    \label{sigma T}
\end{equation}
where $\Phi_0=h/2e$ is the flux quantum. Eq.~\eqref{sigma T} is the
main result of the calculation, which describes the temperature
dependence of $\sigma_I$ for \textit{any} smooth CPR. According to
Eq.~\eqref{sigma T} the power of $2/3$ in the temperature scaling
for $\sigma_I$ is only expected if $I_C(T)=const$. In the SGS
junctions the critical current keeps increasing down to the very low
temperatures (Fig.~\ref{Fig3}c and ~\ref{Fig3}d), due to the
divergence of the normal metal coherence length in graphene, thus
leading to the stronger proximity effect. The solid lines in the
figures are the fits to the SNS junction theory~\cite{ZZ}. The
fitting parameters for the theoretical fits are mean free path
$l_{e}$=$10$-$25\,$nm, which is similar to previously reported
values~\cite{Heersche-Nature07,Miao-Science07,Du-PRB08}, and the
normal resistance $R_{N}$, which is of the same order of magnitude
as the one measured directly.

To further confirm our conclusions we plot $\sigma_I/I^{1/3}_{C}$
versus the temperature as suggested by Eq.~\eqref{sigma T}. The
results are shown in Fig.~\ref{Fig3}e and ~\ref{Fig3}f. Such
critical-current-normalized dispersion obeys the power law with the
power close to $0.6$. For sample 105, the power rests between $0.47$
and $0.62$; for sample 111s these values vary between $0.55$ and
$0.63$, which are close to $2/3$, predicted by the adopted Kurkij\"arvi model.

Another type of scaling which is suggested by Eq.~\eqref{sigma T}
and which can be accessed experimentally is the dependence of the
dispersion on the critical current at \emph{constant} temperature.
Doping-dependent conductivity of graphene provides  a unique
possibility to vary the critical current while keeping the
temperature constant - an experimental ``knob'' which is
inaccessible for other types of junctions.

In Figs.~\ref{Fig3}g and \ref{Fig3}h we present results of such
measurements. We plot the $\ln(\sigma_I)$ versus $\ln(I_{C})$ for
various temperatures (Fig.~\ref{Fig3}g) and scaled  $\ln(\sigma_I)$
versus scaled $\ln(I_{C})$ (Fig.~\ref{Fig3}h). The average value of
the power for sample 111s is 0.34. The power of scaled data for
sample 105 is 0.38, which is shown as the best fit of the data. The
resulting powers are very close to the theoretically expected value
of 1/3.

In summary, we have studied the dispersion of the switching current
distribution in moderately underdamped SGS junctions with clear
hysteretic $I$-$V$ characteristics. A systematic measurements of
both temperature and critical current scaling (at constant $T$) of
the dispersion is performed. The latter study, unavailable in
regular junctions, is made possible by a gate-voltage-tuned
conductivity of graphene. The temperature scaling of the switching
dispersion shows unusual power laws, which is explained
theoretically by taking into account the temperature variation of
the critical current. The critical current scaling of the dispersion
is explained theoretically by combined Stewart-McCumber and
Kurkij\"{a}rvi models, and is applicable for the mesoscopic
junctions with arbitrary current-phase relationships.

The work was supported by ONR grant N000140910689. V.V. was
supported by the Center for Emergent Superconductivity funded by
DOE, under Award No.~DE-AC0298CH1088.

\begin{widetext}
\end{widetext}

\appendix

\section{Supplementary Materials}

\begin{figure}[h!]
  \includegraphics[width=8cm]{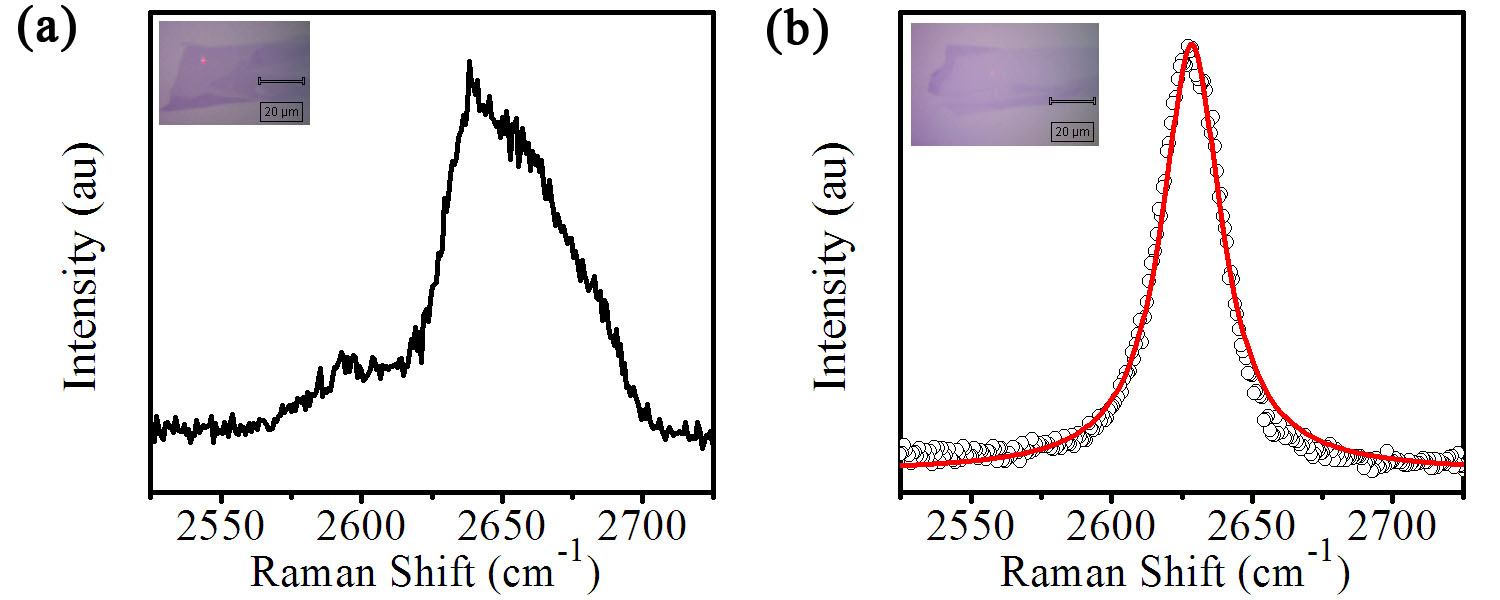}\vskip-.15cm
  \caption{[Color online] (a) Raman spectroscopy for a thick graphite
flake showing a double peak. Inset: optical image of the thick
graphite flake. The red dot shows where the data are recorded. (b)
Raman spectroscopy of a thin graphite flake showing a single peak.
The peak is fit to a Lorentzian curve centered about $2628.5$
cm$^{-1}$ with FWHM of 13 cm$^{-1}$ suggesting that it is graphene.
Inset: optical image of the graphene
flake.}\label{Fig-Raman}\vskip-.15cm
\end{figure}

\emph{Raman spectra and additional sample preparation details}.-- In
Fig.~\ref{Fig-Raman}, the Raman spectroscopy results are plotted for
two different thin graphite flakes. In the thicker flake, the peak
profile is non-Lorentzian and shows a small side peak. In addition,
the FWHM of the best fit Lorenzian is 33 cm$^{-1}$. However, in the
thin flake, the peak profile is Lorentzian and located between
$2600$ and $2700$ cm$^{-1}$. In addition, the FWHM for this sample
is 13 cm$^{-1}$, which is suggestive that the thin flake is
graphene~\cite{Raman3}.

A pair of pseudo-four-probe electrodes are patterned using ebeam
lithography. In a thermal evaporator, a layer of 4nm Pd is
evaporated at a rate of 0.5-1.0 A/s as slower evaporation rates
would lead to a high contact resistance, possibly due to heating of
the graphene by radiation coming from the molten target. Following
the first layer, a layer of 100 nm Pb is evaporated at a rate of
10-30 A/s. We have found that a faster Pb evaporation provides more
uniform films, thus it is desirable. We have used a liquid nitrogen
trap to remove the residual contaminants in the chamber and thus to
increase the quality of the films. After evaporation, the sample is
placed in a hot acetone bath for total 5 minutes for the lift-off
step.  While being in the acetone bath, the sample is sonicated for
10 seconds every other minute. The sample is kept in the second bath
of acetone for 5 minutes before it is rinsed with iso-propanol and
dried. The Pb electrodes are too soft to use wire bonder thus indium
dots have been used to connect the leads to the chip. The sample is
measured in a He$_3$ system. The system is equipped with room
temperature of Pi Filters and low temperature Cu powder filter.

\begin{figure}[b!]
  \includegraphics[width=8cm]{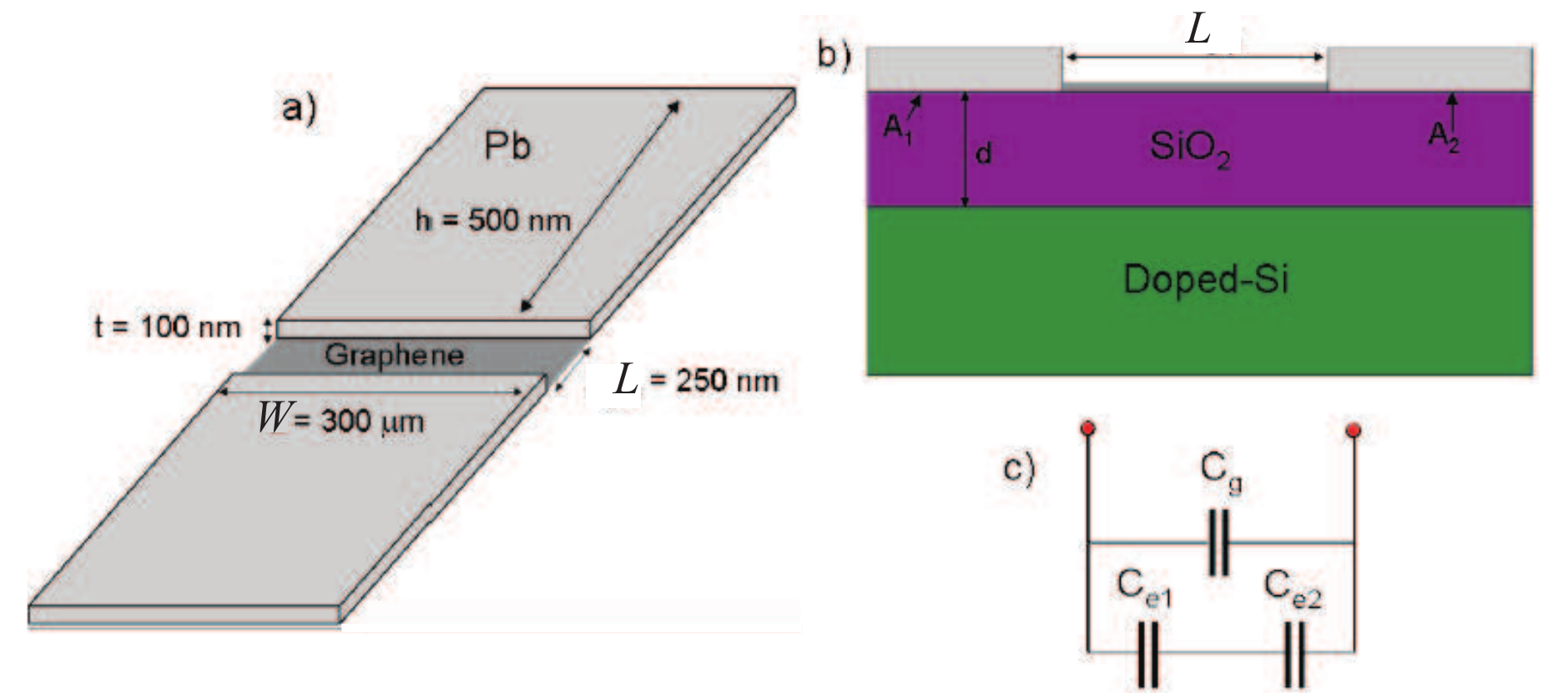}\\
  \caption{[Color online] (a) The capacitance across the graphene, $C_g$, can be calculated
using the cross sectional area of the Pb film and the distance
between the Pb electrodes. (b) The Pb electrodes form capacitors
through the SiO$_2$ with a plate separation of $d$. (c) The
electrical scheme of the capacitance through the graphene and the
two capacitors formed between the electrodes and the gate. The
equivalent capacitance is
$C_g+C_{e1}C_{e2}/(C_{e1}+C_{e2})$.}\label{Fig-C}
\end{figure}

\emph{Remarks on junction capacitance}.-- Total capacitance $C$ is
an important characteristic of the junctions. It defines quality
factor $Q$ which in a way determines whether junction's dynamics is
overdamped or underdamped. In the main text we determined $Q$ from
the experimental values of the critical and retrapping currents,
which together with the junction's normal resistance provided us an
estimate for $C$. An alternative way to determine $C$ is suggested
by considering the geometry of our sample, see Fig.~\ref{Fig-C}. An
estimate of the capacitance across the graphene sheet can be found
with the help of the formula for coplanar
electrodes~\cite{Song-APL70}:
$C_g=\epsilon_r\epsilon_0WK(k)/K(\sqrt{1-k^2})$, which leads us to
the conclusion that the electrode capacitance should be $\sim10$ fF.
Here $\epsilon_0$ is the permittivity of free space, $\epsilon_r$ is
the relative permittivity of silicon $\sim 4$, $W$ is the junction
width $\sim 300\mu$m, and $K(k)$ is the complete elliptical integral
of modulus $k=(1+L/W)^{-1}$; $L$ is the distance between electrodes.
Note, however, that two electrodes have an additional capacitive
coupling through the gate. Thus Fig.~\ref{Fig-C}b gives the proper
geometry and electrical scheme for the calculation of the full
capacitance across the junction. In this geometry, $d$ is $280$nm,
and the area of each electrode (including the presence of a pressed
indium dot to facilitate the electrical connection to the pins on
the chip carrier) is $2.9$ and $2.27$ $\mu$m$^2$. Thus the
capacitance of each electrode to the gate is $36.7$ and $28.7$ pF,
which is calculated from the formula $C_e=\epsilon_r\epsilon_0A/d$,
where $d$ is the distance to the substrate. Using the circuit scheme
as in Fig.~\ref{Fig-C}c, the total capacitance
$C=C_g+C_{e1}C_{e2}/(C_{e1}+C_{e2})$ is found to be $16$pF, which is
in between the values that can be determined from the quality factor
argument. Indeed, for $330$ mK, the quality factor was found to vary
between 3.5 and 4.2 for the entire range of gate voltages from the
values of $I_C$ and $I_{RT}$. Thus the SGS junction (sample 105) is
moderately underdamped as it is larger than 0.84. Since the quality
factor is $Q=w_{p}R_NC$ where $w_{p}=\sqrt{2eI_{C}/\hbar C}$, we
calculate that the capacitance of the sample varies between 12 pF
and 50 pF. This consideration proves consistency of our estimates
between two independent approaches, and confirms the conclusion that
our SGS junction is in the underdamped regime.

\begin{figure}[t!]
  \includegraphics[width=8cm]{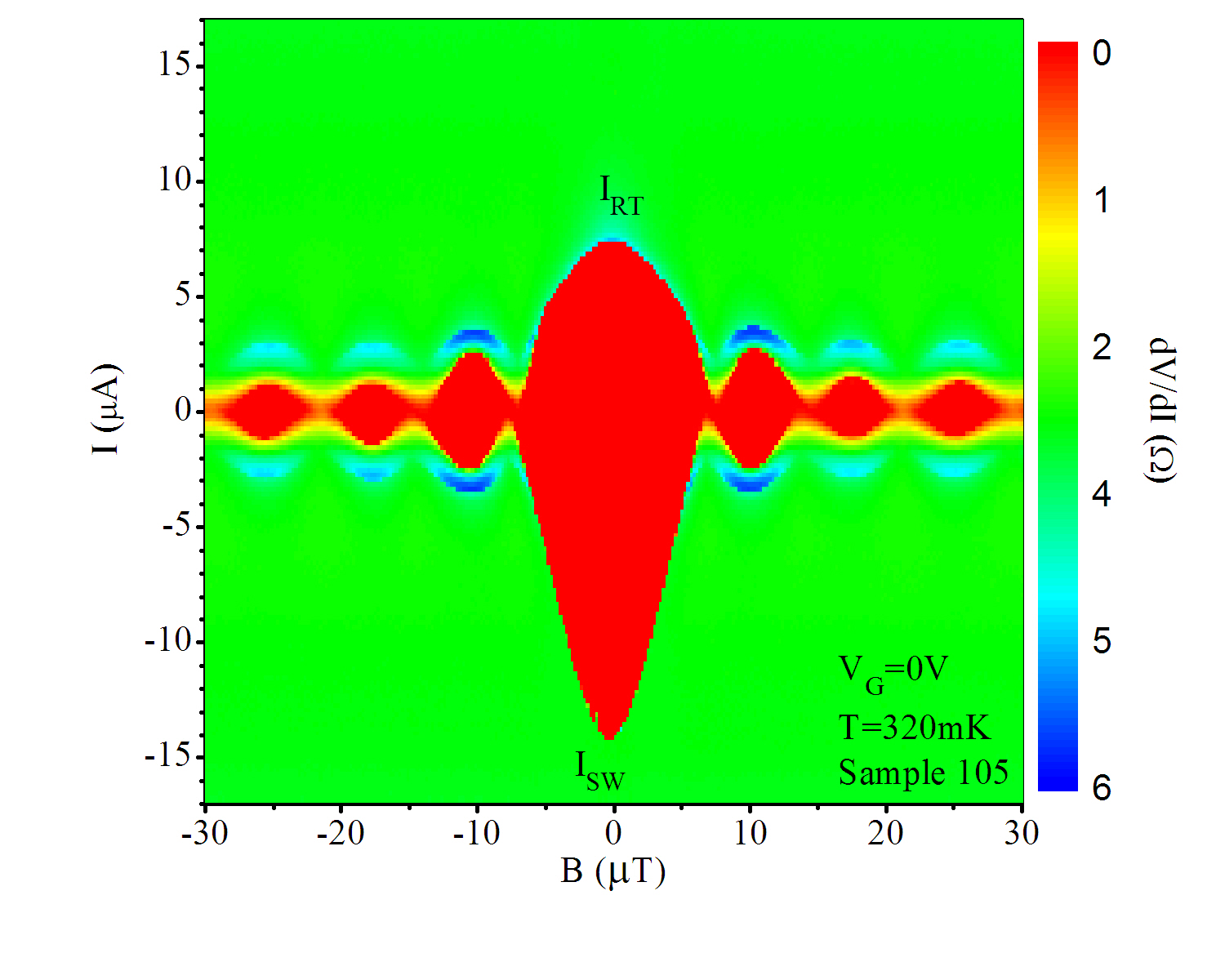}\vskip-.65cm
  \caption{[Color online] A color plot of the differential resistance
  versus bias current and magnetic field. The red color is zero
  differential resistance while the green is the normal state resistance.
  This plot also shows the expected Fraunhofer pattern.}\label{Fig-F}\vskip-.5cm
\end{figure}

\emph{Characterization of graphene proximity junctions}.-- In this
section, the basic characterization of the electronic transport
properties of graphene proximity junctions will be presented. We
focus mainly on the magnetic field dependence of the supercurrent,
which demonstrates high quality JJ in our graphene based devise, and
on the gate voltage dependence of resistance and switching current,
which traces properties of the graphene, associated with the Dirac
spectrum. This discussion complements presentation given in the main
text.

The magnetic field dependence of the switching current can be
measured and compared to Josephson junction theory. In an extended
JJ, the maximum (or critical) current as the function of applied
magnetic flux can be fit to a Fraunhofer function:
\begin{equation}
\frac{I(\Phi)}{I(0)}=\left|\frac{\sin(\pi\Phi/\Phi_0)}{\pi\Phi/\Phi_0}\right|\,,
\end{equation}
where $\Phi$ is the magnetic flux through the junction which is
given by $\Phi=BW(L+\lambda/2)$, $B$ is the magnetic field, and
$\Phi_0=h/2e$ is the magnetic flux quantum. The effective area is
larger than just $LW$ due to the magnetic field penetration into the
electrodes by a distance equal to the penetration depth $\lambda$.

The fit of our data to the Fraunhofer pattern can be seen in
Fig.~\ref{Fig1}d of the main text, where a good fit is obtained
using an area that is estimated from the junction area of
$\sim~10^{-10}$ m$^2$, which results in a magnetic field period of
$\sim 20$ mT. A vertical shift of $2.5$ mA was also used to obtain
the fit to account for the supercurrent observed even at the Dirac
point. The magnetic field period obtained from the experimental data
is somewhat small at about $7$ mT. This discrepancy between the
fitted and measured value can be traced to an understanding of the
effective area of the junction $W(L+\lambda/2)$. Using a penetration
depth of $120$ nm results in an effect area of $\sim 2\times
10^{-10}$ m$^2$ and a magnetic field period of $9$ mT. If the area
of the entire electrode plus junction were used, the magnetic field
period would be $\sim 7.8$ mT. Thus, to fit to the Fraunhofer period
the penetration depth in our samples should be larger than $120$ nm
in order to increase the effective area of the junction. The
differential resistance of the junction can also be measured as a
function of bias current and magnetic field, see Fig.~\ref{Fig-F}.
Again, the Fraunhofer pattern is observed, with a slightly reduced
value of the switching current even at a applied gate voltage of
$50$ V due to a shift of the Dirac point after the thermal cycling.

\begin{figure}[b!]
  \includegraphics[width=8cm]{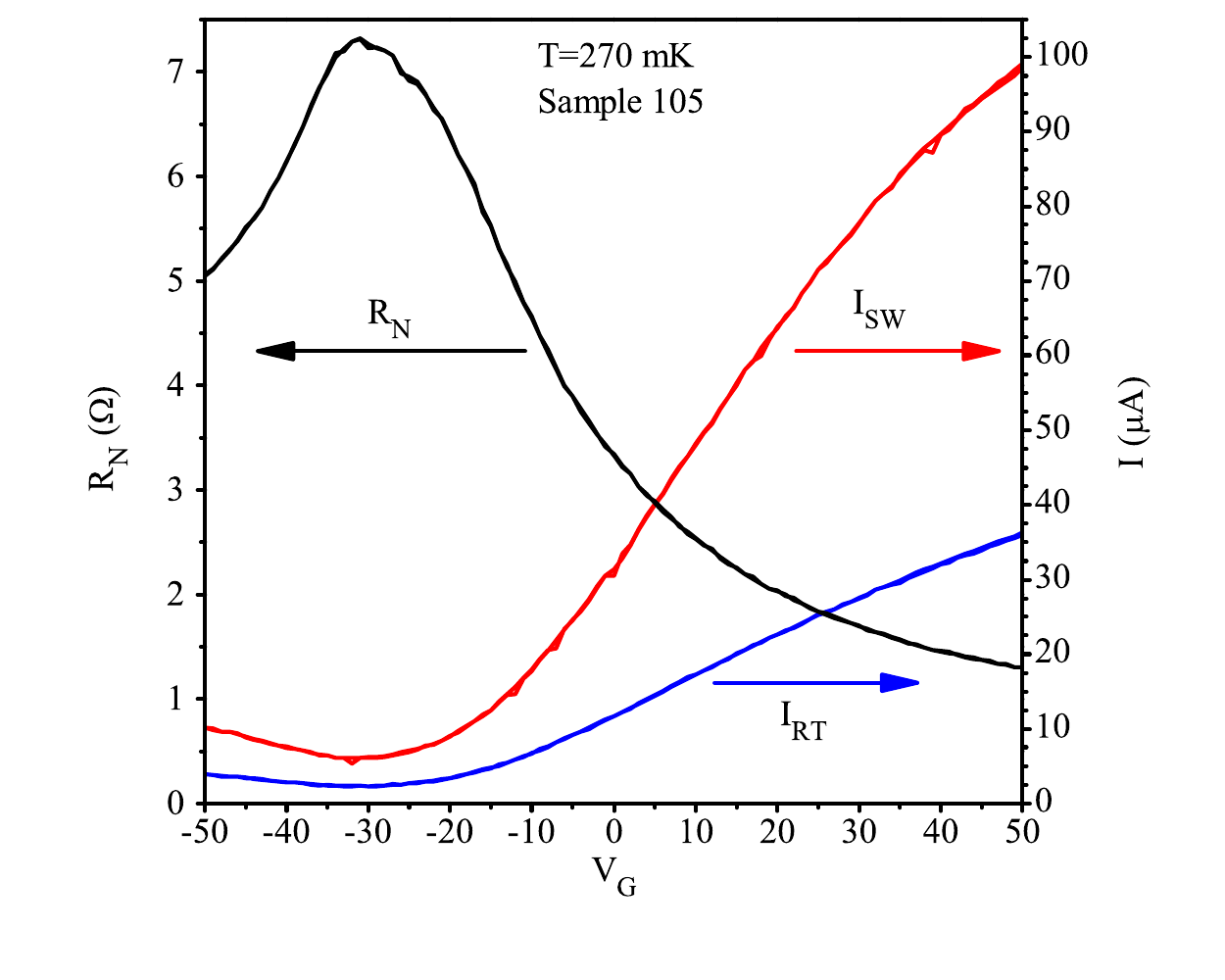}\vskip-.5cm
  \caption{[Color online] (a) The normal resistance,
  the retrapping current, and the switching current
  shown as a function of gate voltage.
  As the gate voltage is swept from the Dirac point,
  the switching current and retrapping current increase
  however resistance of the sample decreases.
 }\label{Fig-dVdI}\vskip-.5cm
\end{figure}

The switching current and resistance as a function of the gate
voltage are displayed in Fig.~\ref{Fig-dVdI}. As the gate voltage is
swept away from the Dirac point, the total carrier concentration
increases and thus, the switching current is expected to increase as
well, which is confirmed in Fig.~\ref{Fig-dVdI}a. Similarly, the
retrapping current increases as well. In addition, the resistance of
the sample decreases as the carrier concentration increases. In
Fig.~\ref{Fig-dVdI}a the Dirac point is observed to be at $\sim 21$
V and above this value the Fermi energy is shifted to the conduction
band and thus the charge carriers are primarily electrons while
below $21$ V, the Fermi energy is shifted to the valence band and
the charge carriers are mostly holes. Thus, this figure confirms
that our junctions are capable of carrying a bipolar supercurrent.
It is also clear that the gate voltage can be used to tune the
critical current of our graphene proximity junctions, which can be
used as a tool to study the escape dynamics.

\emph{Current-phase relation in graphene and $I_C(T)$}.-- Although
detailed information about CPR in the SGS proximity junction was not
needed for our analysis of current switching statistics,
nevertheless, we provide such information for completeness. The
equilibrium Josephson current can be found from the thermodynamic
relation
\begin{equation}\label{I-def}
I_S(\phi)=\frac{2e}{\hbar}\frac{dF}{d\phi}
\end{equation}
by knowing free energy $F$ as a function of the phase difference
across the junction. Bardeen \textit{et. al.}~\cite{Bardeen-PR69}
derived general expression for $F$ with the help of Bogolubov-de
Gennes (BdG) equations, which reads
\begin{equation}\label{F}
F=-2T\sum_n\ln[2\cosh(\varepsilon_n/2T)]+\int d^dr |\Delta|^2/V\,.
\end{equation}
This result is a generalization of that used in BCS model for the
case of $\Delta=const$. The sum over $n$ is just what would be
obtained for the free energy for an assembly of independent fermions
with energies $\varepsilon_n$. As in the Hartree-Fock approximation,
this counts the interaction energy twice. The next term, $\int
d^dr|\Delta|^2/V$, the negative of the interaction energy, corrects
for this double counting. Since the bulk energy $\propto |\Delta|^2$
is independent of the phase $\phi$ one obtains from Eq.~\eqref{F}
after differentiation
\begin{equation}\label{I}
I_S=-\frac{4e}{\hbar}\!\!\sum_{\varepsilon_n<\Delta}\!\!
\frac{d\varepsilon_n}{d\phi}\tanh\frac{\varepsilon_n}{2T}
-\frac{8eT}{\hbar}\!\!\int^{\infty}_{\Delta}\!\!
d\varepsilon\ln\left[2\cosh\frac{\varepsilon_n}{2T}\right]\frac{d\rho}{d\phi}
\end{equation}
where we have rewritten sum over $n$ as a sum over the discrete
positive eigenvalues $\varepsilon_n(\phi)$ $(n = 1, 2, . . .)$ of
BdG equations, and an integration over the continuous spectrum with
density of states $\rho(\varepsilon,\phi)$. The additional factor of
$2$ in the above formula, as compared to the conventional
expression, accounts for two valleys in graphene. Although above the
gap states, $\varepsilon>\Delta$, do contribute to the Josephson
current~\cite{AL} their role, in fact, is subleading as compared to
the contribution coming from the Andreev bounds states with energies
below the gap $\varepsilon_n<\Delta$. One thus only need to consider
the first term in Eq.~\eqref{I}. In the context of graphene the
spectrum of Andreev levels can be found from the BdG equations in
the following form~\cite{Titov-PRB06}
\begin{equation}\label{BdG}
[(ihv\bm{\sigma}\cdot\bm{\partial}-\mu)\otimes\tau_z]\vec{\psi}=\varepsilon\vec{\psi}\,,
\end{equation}
where $\vec{\psi}^T=(\psi_e,\psi_h)$ is electron(hole) wave
functions combined in the vector, dot product is defined in the
conventional way
$\bm{\sigma}\cdot\bm{\partial}=\sigma_x\partial_x+\sigma_y\partial_y$,
symbol $\otimes$ stands for the direct product between matrices,
which operate in the isospin ($\sigma$) and electron-hole ($\tau$)
spaces. Chemical potential $\mu$ is measured with respect to Dirac
point, so that $\mu=0$ corresponds to undoped graphene.
Electron-like and hole-like wave functions are related to each other
at the SG interfaces $\bm{r}_\pm=(\pm L/2,y)$ via specular
reflection Andreev processes. Mathematically this can be represented
as follows~\cite{Titov-PRB06}
\begin{subequations}\label{BC}
\begin{equation}
\psi_h(\bm{r}_-)=U(\varepsilon)\psi_e(\bm{r}_-)\,,\quad
\psi_h(\bm{r}_+)=U^{-1}(\varepsilon)\psi_e(\bm{r}_+)\,,
\end{equation}
\begin{equation}
U(\varepsilon)=\exp(-i\phi/2+i\beta\sigma_x)\,,\quad
\beta=\arccos(\varepsilon/\Delta)\,.
\end{equation}
\end{subequations}
 Assuming hard wall
boundary conditions in the $y$-direction $k_y$-component of particle
wave vector becomes quantized $k_y\equiv q_n=(n+1/2)\pi/W$.
Performing then Fourier transform in Eq.~\eqref{BdG}, matching
electron/hole plane waves at both interfaces with the help of
boundary conditions \eqref{BC} and setting determinant of the
corresponding matrix equation to zero one finds
\begin{equation}
\cos\phi=\left(\cos^2\chi+\frac{\sin^2\chi}{\cos^2\gamma}\right)\cos2\beta-\sin^2\chi\tan^2\gamma
\end{equation}
which determines dispersion relation for the Andreev levels, where
$\chi=kL$, $k=(\mu/\hbar v)\cos\gamma$, $\gamma=\arcsin(\hbar
vq_n/\mu)$. The last equation can be resolved analytically for
$\varepsilon$ as the function of the channel index $n$ and
superconducting phase difference $\phi$ in the form
\begin{equation}\label{AL}
\varepsilon_n(\phi)=\Delta\sqrt{1-|t_n|^2\sin^2(\phi/2)}\,,
\end{equation}
where $|t_n|^2$ has meaning of the transmission coefficient in the
$n^{th}$ transversal channel, which is given explicitly by
\begin{equation}
|t_n|^2=\frac{\kappa^2_n}{\kappa^2_n\cos(\kappa_nL)+(\mu/\hbar
v)^2\sin^2(\kappa_nL)}\,,
\end{equation}
with $\kappa_n=\sqrt{(\mu/\hbar v)^2-q^2_n}$. Note that at the Dirac
point, $\mu\to0$, all channels are evanescent, since $\kappa_n\to
iq_n$ and thus
\begin{equation}\label{t}
|t_n|^2=\frac{1}{\cosh^2(q_nL)}\,.
\end{equation}
Having determined the spectrum of energy states below the gap we can
return to Eq.~\eqref{I} and find Josephson current in the
form~\cite{Cserti-PRB10}
\begin{equation}
I_S(\phi)=\frac{e\Delta^2}{\hbar}\sum_{\varepsilon_n<\Delta}
\frac{|t_n|^2\sin\phi}{\varepsilon_n(\phi)}\tanh\frac{\varepsilon_n(\phi)}{2T}
\end{equation}
where $\varepsilon_n(\phi)$ should be taken from Eq.~\eqref{AL}. In
the vicinity of the neutrality point, $\mu\ll E_{Th}$, where
$E_{Th}=\hbar v/L$ is ballistic Thouless energy, to the good
approximation one can use Eq.~\eqref{t} for the transmission
coefficient. Furthermore, if the aspect ratio of graphene sheet is
such that $W\gg L$ then summation over discrete $n$ can be replaced
by the integration $\sum_n\to\frac{W}{\pi L}\int^{\infty}_{0}dx$
with $q_nL\to x$. Introducing also dimensionless variable
$z=\sqrt{1-\sin^2(\phi/2)/\cosh^2x}$ one arrives at
\begin{equation}
I_S(\phi)=\frac{2e\Delta W}{\pi\hbar
L}\int^{1}_{\cos(\phi/2)}dz\frac{\cos(\phi/2)
\tanh\frac{z\Delta}{2T}}{\sqrt{z^2-\cos^2(\phi/2)}}\,.
\end{equation}
There is no close analytical expression for this integral, except
for the zero temperature limit, when $\tanh\frac{z\Delta}{2T}\to1$.
In that case~\cite{Titov-PRB06,KO}
\begin{equation}\label{Ic-SGS}
I_S(\phi)=\frac{e\Delta W}{\pi\hbar
L}\cos(\phi/2)\ln\left(\frac{1+\sin(\phi/2)}{1-\sin(\phi/2)}\right)\,,
\end{equation}
which coincides with the result of Kulik-Omelyanchuk for the case of
disordered SNS junction~\cite{KO}. This is rather peculiar result
since calculation was done for the manifestly ballistic limit of
graphene. From the CPR we can restore free energy barrier for phase
slips and thus hight of the Gibbs potential (see Eq.~3 in the main
text):
\begin{equation}\label{Gc-SGS}
G_C=\frac{2c^{3/2}_{1}}{\pi\sqrt{c_2}}\frac{\Delta
W}{L}=\frac{4}{\pi^2}\sqrt{\frac{c_1}{c_2}}eI_CR_N
\end{equation}
where we have used $I_C=c_1\frac{e\Delta W}{\pi\hbar L}$,
$|\partial^2_\phi I_S|_{\phi=\phi_C}=c_2\frac{e\Delta W}{\pi\hbar
L}$, $R^{-1}_N=4e^2W/\pi hL$ with $c_1=1.33, c_2=1.08$, and
$\phi_C=1.97$ which corresponds to the maximum of $I_S(\phi)$. For
completeness we mention that in the longer junctions, $L\gg W$, all
transmissions are exponentially suppressed, $|t_n|^2\approx
e^{-q_nL}$, such that summation reduced to the geometrical
progression to the leading order in $|t_n|^2$, with the result
\begin{equation}
I_S(\phi)=\frac{e\Delta}{\hbar}\tanh\left(\frac{\Delta}{2T}\right)e^{-\pi
L/W}\sin\phi\,.
\end{equation}
Note that for our samples the aspect ratio is at least $W/L=40$ so
that Eqs.~\eqref{Ic-SGS}--\eqref{Gc-SGS} should apply near the
neutrality point.

At the relatively high doping when $E_{Th}\ll \mu\ll \Delta$ full
expression for $t_n$ is needed for the calculation of $I_S(\phi)$.
Although there is no close analytical expression for the CPR in this
case, critical Josephson current can be estimates as
\begin{equation}
I_C\simeq N_{ch}\frac{e\Delta}{\hbar}\,,\quad N_{ch}=\frac{\mu
W}{\pi\hbar v}\,,
\end{equation}
where $N_{ch}$ has meaning of the number of propagating transversal
channels. All above considerations apply, strictly speaking, to the
short junctions $L\ll\xi$ in the sense of the proximity effect,
where $\xi$ is superconducting coherence length. Note, however, that
our SGS devices are rather at the crossover between long and short
limits with the typical ratio $L/\xi\sim2$.

Surprisingly, we do not find solid experimental evidence in support
of ballistic transport in the SGS proximity junctions. All aspects
of our data, and in particular temperature and gate voltage
dependence of the critical current, are in fact in good quantitative
agreement with the predictions of \textit{diffusive} SNS junction
model~\cite{AL,ZZ-app,Dubos-PRB01-App}. Specifically we use
\begin{equation}\label{Ic-SNS}
eI_CR_N=64\pi
T\sum^{\infty}_{n=0}\frac{\Delta^2(L/L_\varepsilon)\exp(-L/L_\varepsilon)}
{[\varepsilon_n+E_n+\sqrt{2(E^2_n+\varepsilon_nE_n)}]^2}
\end{equation}
to fit the experimental data (see Fig.~3c and 3d of the main text).
In Eq.~\eqref{Ic-SNS} $\varepsilon_n=(2n+1)\pi T$,
$E_n=\sqrt{\Delta^2+\varepsilon^2_n}$ and
$L_\varepsilon=\sqrt{D/2\varepsilon_n}$. In theory
Eq.~\eqref{Ic-SNS} applies at $T\gg E_{Th}$, where $E_{Th}=D/L^2$ is
diffusive Thouless energy and $D=vl_e/2$ is diffusion coefficient.
For the typical parameters of our samples $v\sim10^8$m/s,
$l_e\sim10$ nm, $L\sim 300$ nm one finds $E_{Th}\sim0.1$ K which
corresponds to $T\gtrsim E_{Th}$ for the working temperature regime
$T=0.3$ K. Note also that as shown in the extensive study of
Ref.~\cite{Dubos-PRB01-App}, Eq.~\eqref{Ic-SNS} works well in the
rather wide temperature range. At the lowest temperatures, $T\ll
E_{Th}$, critical current saturates with the exponential
accuracy~\cite{AL,Dubos-PRB01-App}
\begin{equation}
eI_CR_N\simeq aE_{Th}[1-b\exp(-cE_{Th}/T)]\,,
\end{equation}
where $a=10.8$, $b=1.3$, and $c=3.4$. Finally, one should note that
product $eI_CR_N$ (up to a numerical factor of order one) sets the
magnitude of the Gibbs barrier for the phase slips. For example, at
$T\sim 0.4$ K in the case of sample 105 taking the corresponding
values $I_C\sim10$ $\mu A$ and $R_N\sim 10$ $\Omega$ one estimates
$G_C\sim eI_CR_N\sim 1$ K, such that $G_C\gg T$ in agreement with
our earlier discussions in the main text.


\vspace{-.4cm}

\begin{thebibliography}{00}

\bibitem{Novoselov-05}
K.~S.~Novoselov, \textit{et al.}, Nature \textbf{438}, 197 (2005).

\bibitem{Zhang-05}
Y.~Zhang, \textit{et al.}, Nature \textbf{438}, 201 (2005).

\bibitem{Geim-07}
A.~K.~Geim and K.~S.~Novoselov, Nat. Mater. \textbf{6}, 183 (2007).

\bibitem{Geim-09}
A.~K.~Geim, Science \textbf{324}, 1530 (2009).

\bibitem{Castro-Neto-09}
A.~H.~Castro Neto, \textit{et al.}, Rev. Mod. Phys. \textbf{81}, 109
(2009).

\bibitem{Beenakker}
C.~W.~J.~Beenakker, Phys. Rev. Lett. \textbf{97}, 067007 (2006);
Rev. Mod. Phys. \textbf{80}, 1337 (2008).

\bibitem{Meissner-PRL59}
H.~Meissner, Phys. Rev. Lett. \textbf{2}, 458 (1959).

\bibitem{Heersche-Nature07}
H.~B.~Heersche, \textit{et al.}, Nature \textbf{446}, 56 (2007).

\bibitem{Miao-Science07}
F.~Miao, \textit{et al.}, Science \textbf{317}, 1530 (2007).

\bibitem{Du-PRB08}
X.~Du, I.~Skachko, and E.~Y.~Andrei, Phys. Rev. B \textbf{77},
184507 (2008).

\bibitem{Ojeda-PRB09}
C.~M.~Ojeda-Aristizabal, \textit{et al.}, Phys. Rev. B \textbf{79},
165436 (2009).

\bibitem{Jeong-PRB11}
D.~Jeong, \textit{et al.}, Phys. Rev. B \textbf{83}, 094503 (2011).

\bibitem{Borzenets}
I.~V.~Borzenets, \textit{et al.}, Phys. Rev. Lett. \textbf{107},
137005 (2011).

\bibitem{Lee-arXiv11}
G.~H.~Lee, \textit{et al.}, Phys. Rev. Lett. \textbf{107}, 146605
(2011) .

\bibitem{MDC}
J.~M.~Martinis, M.~H.~Devoret, and J.~Clarke, Phys. Rev. B
\textbf{35}, 4682 (1987).

\bibitem{Sahu}
M.~Sahu, \textit{et al.}, Nature Physics \textbf{5}, 503  (2009);
P.~Li, \textit{et al.}, Phys. Rev. Lett. \textbf{107}, 137004
(2011).

\bibitem{Yu}
H.~F.~Yu, \textit{et al.}, Phys. Rev. Lett. \textbf{107}, 067004
(2011).

\bibitem{Warburton2009}
P.A.~Warburton, \textit{et al.}, Phys. Rev. Lett. \textbf{103},
217002 (2009).

\bibitem{FD}
T.~A.~Fulton and L.~N.~Dunkleberger, Phys. Rev. B \textbf{9}, 4760
(1974).

\bibitem{JK}
J.~Kurkij\"{a}rvi, Phys. Rev. B \textbf{6}, 832 (1972).

\bibitem{Raman2}
D.~Graf \textit{et al.}, Nano Letters \textbf{7}, 238 (2007).

\bibitem{Tinkham}
M.~Tinkahm, \textit{Introduction to Superconductivity}, 2d ed.
(McGraw-Hill Inc. 1996).

\bibitem{SM}
W.~C.~Stewart, Appl. Phys. Lett. \textbf{12}, 277 (1968);
D.~E.~McCumber, J. Appl. Phys. \textbf{39}, 3133 (1968).

\bibitem{Kramers}
H.A.~Kramers, Physica {\bf 7}, 284 (1940).

\bibitem{ZZ}
A.~D.~Zaikin and G.~F.~Zharkov, Sov. J. Low. Temp. Phys.
\textbf{7(3)}, 184 (1981); P.~Dubos \textit{et al.}, Phys. Rev. B
\textbf{63}, 064502 (2001).

\end{thebibliography}

\vspace{-.45cm}

\begin{thebibliography}{00}

\bibitem{Raman3}
D.~Graf \textit{et. al.}, Nano Letters \textbf{7}, 238 (2007).

\bibitem{Song-APL70}
Y.~Song, J. Appl. Phys. \textbf{47}, 2651 (1976).

\bibitem{Bardeen-PR69}
J.~Bardeen, R.~K\"{u}mmel, A.~E.~Jacobs, and L.~Tewordt, Phys.
Rev. \textbf{187}, 556 (1969).

\bibitem{AL}
A.~Levchenko, A.~Kamenev, and L.~Glazman, Phys. Rev. B \textbf{74},
212509 (2006); Phys. Rev. B \textbf{77}, 180503(R) (2008).

\bibitem{Titov-PRB06}
M.~Titov and C.~W.~J.~Beenakker, Phys. Rev. B \textbf{74}, 041401(R)
(2006).

\bibitem{Cserti-PRB10}
I.~Hagym\'{a}si, A.~Korm\'{a}nyos, and J.~Cserti, Phys. Rev. B
\textbf{82}, 134516 (2010).

\bibitem{KO}
I.~O.~Kulik and A.~N.~Omelyanchuk, JETP Lett. \textbf{21}, 96
(1975).

\bibitem{ZZ-app}
A.~D.~Zaikin and G.~F.~Zharkov, Sov. J. Low. Temp. Phys.
\textbf{7(3)}, 184 (1981).

\bibitem{Dubos-PRB01-App}
P.~Dubos \textit{et. al.}, Phys. Rev. B \textbf{63}, 064502 (2001).

\end{thebibliography}
\end{document}